\documentstyle[aps,epsfig,twocolumn]{revtex}
\title{Kondo effect and singlet-triplet splitting
in coupled quantum dots in a magnetic field}
\author{Vitaly~N.~Golovach and  Daniel Loss}
\address{ Department of Physics and Astronomy, University of Basel,
Klingelbergstrasse 82, CH-4056 Basel, Switzerland}

\date{\today}

\begin{document}
\twocolumn[\hsize\textwidth\columnwidth\hsize\csname
@twocolumnfalse\endcsname
\maketitle

\begin{abstract}
We study two tunnel-coupled quantum dots each with a spin 1/2 and
attached  to leads in the 
Coulomb blockade regime. We study the interplay between 
Kondo correlations and the singlet-triplet exchange splitting $K$
between the two spins.
We calculate the cotunneling current  with
elastic and inelastic contributions
and its renormalization due to Kondo correlations,  away  and at the
degeneracy point $K=0$. 
We show that these Kondo correlations
induce pronounced peaks in the conductance as function
of magnetic
field $B$, inter-dot coupling $t_0$, and temperature. 
Moreover, the long-range part of the Coulomb interaction becomes
visibile due to Kondo correlations resulting in an additional peak 
in the conductance vs $t_0$ with a strong $B$-field dependence.
These conductance peaks thus provide direct experimental access to  $K$, 
and thus to a crucial control parameter for spin-based qubits and entanglement.
\end{abstract}\vspace{-30pt}


\vskip2pc] \narrowtext

Semiconductor quantum dots  have attracted much interest over the 
years due to their rich  and reproducible transport properties in the Coulomb
blockade (CB) regime, where the number of electrons on the dot is
quantized due to charging effects
\cite{Kouwenhoven}. More
recently, attention has focused on the spin of the electron in such
nanostructures, with the spin introducing new correlation effects such as
Kondo behavior\cite{Raikh,GG,Cronenwett,Schmid,Cobden}, and its interplay 
with spin exchange interaction in 
single\cite{Sasaki,Glazman,Nazarov,Hofstetter} 
and double dots \cite{Eto,Izumida}. 
On the other hand, it has been pointed out that the spin, 
confined to a quantum dot or atom, is a suitable candidate for 
quantum information processing\cite{Loss97}, due to unusually long 
decoherence time of spin\cite{Kikkawa}. A crucial element in such 
spin-based quantum computing schemes is the Heisenberg exchange 
interaction  $K$ (singlet-triplet splitting) between spins of adjacent
dots, being controlled via the  interdot tunneling\cite{Loss97}.
Thus, the primary goal is to achieve control over $K$ which
then allows one to generate deterministic
entanglement of spins. Using a
Hund-Mulliken (HM) approach to describe a realistic double dot system (DD) it
has been shown\cite{Loss_99} that $K$ is very sensitive to long range
Coulomb interaction as well as to magnetic fields by which
a singlet-triplet crossing can be tuned. 
Motivated by this we
study here transport and Kondo effects in such a realistic DD
system within the HM approach, thereby going 
beyond short-range on-site models used so far to describe Kondo effects.
In particular, we calculate the current through the DD
via a Schrieffer-Wolff transformation and 
via a systematic
cotunneling calculation including elastic and inelastic contributions. Using a
perturbative renormalization  group (RG) approach we show that
the conductance in the cotunneling regime shows pronounced peaks induced
by Kondo correlations and long range Coulomb
interactions as function of temperature, inter-dot coupling,
magnetic fields, and bias. Such  Kondo enhanced peaks in
the conductance
thus provide direct experimental access to singlet/triplet
states and their exchange splitting $K$--the quantities of
crucial importance for spin-based qubit schemes.

We consider a DD system 
consisting of two lateral quantum dots tunnel-coupled  to
metallic leads, in the presence
of a perpendicular magnetic field $B$, see Fig.~1.
At low temperatures $T$, the conductance $G$ of the DD as a
function of the gate voltage $V_g$ shows sharp doublets of sequential
tunneling peaks separated by CB valleys (cotunneling regime)
\cite{2dots_exper}. In the middle between such peaks, the
number of electrons in the DD is even, $M=2N$ (assuming similar
dots).
We assume that $N-1$ electrons of each dot form a closed shell
[with $N-1$ even] and thus ignore them.
The remaining two electrons in the DD can be described by the Hamiltonian
\begin{equation}\label{H_dot}
H_d=\sum_{i=1,2}\left[\frac{1}{2m}\left({\bf p}_i-
\frac{e}{c}{\bf A}({\bf r}_i)\right)^2+
W({\bf r}_i)\right]+C,
\end{equation}
where $C=e^2/\kappa|\bf{r}_1-\bf{r}_2|$ is the Coulomb interaction, with
charge $e$ and dielectric constant $\kappa$  ($=13.1$ for GaAs).
As usual in GaAs, the Zeeman interaction will be neglected.
\begin{figure}
\narrowtext {\epsfxsize=6.cm
\centerline{\epsfbox{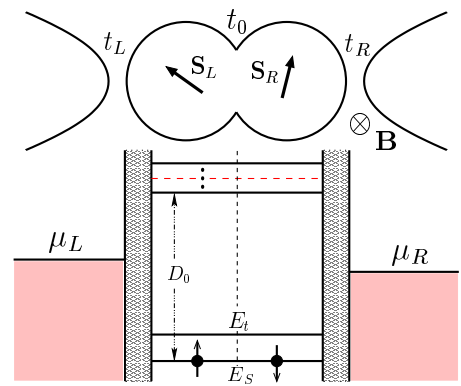}}} \caption{Double-dot 
system containing two electrons and being coupled in series to two
metallic leads at chemical potentials $\mu_R$ and $\mu_L$ with
bias $\Delta\mu=\mu_L-\mu_R$. The electron spins ${\bf S}_L$,
${\bf S}_R$ interact via the exchange interaction $K=E_t-E_S$,
where $E_{t,S}$ is the triplet/singlet energy.} \label{fig1}
\end{figure}
\hspace{-13pt}
However, the orbital effect of the magnetic field
$\bf{B}=\bf{\nabla}\times\bf{A}$ is important, it allows us to tune
the singlet-triplet transition in the isolated DD, provided, however,
there is  long-range Coulomb interaction\cite{Loss_99}.   
Within the low energy sector, a realistic confinement
potential for the DD is given by \cite{Loss_99}
$W({\bf r})=(m\omega_0^2/2)[(x^2-a^2)^2/4a^2+y^2]$,
which separates at $a\gg\sqrt{\hbar/m\omega_0}$ into two
harmonic wells of frequency $\omega_0$, located at $x=\pm a$,
with, typically in GaAs, $h\omega_0\simeq 3\mbox{ meV}$\cite{Kouwenhoven}.
Within the HM approach we express the
two-particle states of the DD\cite{Loss_99} in terms of 
symmetric (for spin singlet) and antisymmetric (for spin triplet) combinations of the Fock-Darwin states, given by 
$\varphi_{\pm a}(x,y)=\exp[{-((x\mp a)^2+y^2)/2\lambda^2\mp iya/2l^2}]/\lambda\sqrt{\pi}$,
where $l=\sqrt{\hbar c/|e|B}$ and $\lambda=\sqrt{\hbar/m\omega}$, and
$\omega=\sqrt{\omega_0^2+\omega_L^2}$, where  $\omega_L=|e|B/2mc$.
We introduce the operators $d_{\pm,\sigma}^{\dag}(d_{\pm,\sigma})$, 
which create (annihilate) the DD states
$\psi_{\pm,\sigma}=\chi_{\sigma}(\varphi_{-a}\pm\varphi_{+a})/\sqrt{2(1\pm S)}$, where $\chi_{\sigma}$ is the spinor and
$S=\langle\varphi_{\pm a}|\varphi_{\mp a}\rangle$ the overlap integral.
The singlet and triplet states then become,
\begin{eqnarray}\label{states}
&&|00\rangle=\frac{1}{\sqrt{1+\phi^2}}(d_{+\uparrow}^{\dag}
d_{+\downarrow}^{\dag}-\phi d_{-\uparrow}^{\dag}
d_{-\downarrow}^{\dag})|0\rangle\;,
\nonumber\\
&&|11\rangle=d_{-\uparrow}^{\dag}d_{+\uparrow}^{\dag}|0\rangle\;,\;\;\;\;
|1-1\rangle=d_{-\downarrow}^{\dag}d_{+\downarrow}^{\dag}|0\rangle\;,\\
&&|10\rangle=\frac{1}{\sqrt{2}}(d_{-\uparrow}^{\dag}
d_{+\downarrow}^{\dag}+d_{-\downarrow}^{\dag}
d_{+\uparrow}^{\dag})|0\rangle\nonumber\;,
\end{eqnarray}
where $|0\rangle$ is the DD ground state with $M=2N-2$, and
\begin{equation}
\phi=\sqrt{1+\left(\frac{4t_H}{U_H}\right)^2}-\frac{4t_H}{U_H}\;,
\label{phi}
\end{equation}
where $t_H=t_0+t_C$ is the extended inter-dot tunneling amplitude
with $t_0$ its bare value ($C=0$) and $t_C(t_0)$ the part that gets 
renormalized by the Coulomb interaction $C$.
Similarly, $U_H$ is the extended
on-site Coulomb repulsion which decreases with decreasing $a$\cite{Loss_99}.
Note that for $t_0=0$ we have $\phi=1$, while $\phi<1$
occurs due to double occupancies in
the dots, and $\phi\to 0$ for $C\to 0$.
The DD is connected  to two Fermi liquid leads
$\alpha=L,\;R$ (Fig.~1), described by
$H_l=\sum_{\alpha k\sigma}\varepsilon_kc_{\alpha k\sigma}^{\dag}c_{\alpha k\sigma}$,
where $c_{\alpha k\sigma}^{\dag}$ creates an electron-state with
momentum $k$ and spin $\sigma$ in lead $\alpha$.
The tunnel coupling from lead $\alpha$ to the nearest dot is parametrized 
by the amplitude $t_{\alpha}$, and thus 
the  amplitude to tunnel from lead $\alpha$ into the DD states
$\psi_{n\sigma}$ is then given by 
$t_{L,\pm}=t_L/\sqrt{2(1\pm S)}$, $t_{R,\pm}=\pm t_R/\sqrt{2(1\pm S)}$.
We use the tunneling Hamiltonian 
$H_T=\sum_{n\alpha k\sigma}(t_{\alpha n}
c_{\alpha k\sigma}^{\dag}d_{n\sigma}+\mbox{h.c.})$ to perform a 
Schrieffer-Wolff transformation\cite{SW}, and arrive at the  
cotunneling part $H_{\mbox{\em cot}}=H_T[(i0^+-\hat{L}_0)^{-1}H_{T}]$, with
Liouvillean 
${L}_0A\equiv[H_0,A]$, 
where $H_0=H_{d}+H_l$.

Next we concentrate on the states (\ref{states}) and 
project out the higher lying energy levels by means
of the mapping~\cite{Glazman}
\begin{eqnarray}\label{mapping}
&&\sum_{\sigma\sigma'}d_{n\sigma}^{\dag}
\vec{\bf\sigma}_{\sigma\sigma'}d_{n'\sigma'}={\bf S}_+\delta_{nn'}+
\left(\frac{\phi_+}{2}{\bf
S}_-+in\phi_-{\bf T}\right)\delta_{-nn'},\nonumber
\end{eqnarray}
\vspace{-12pt}
\begin{eqnarray}
&&\sum_{\sigma}d_{n\sigma}^{\dag}d_{n'\sigma}
=\delta_{nn'}\left[1-\frac{n}{2}\phi_+\phi_{-}\left({\bf
S}_L\cdotp{\bf S}_R-\frac{1}{4}\right)\right],\;\;\;\;
\end{eqnarray}
where ${\bf S}_{\pm}={\bf S}_L\pm{\bf S}_R$, ${\bf T}={\bf
S}_L\times{\bf S}_R$, and
$\phi_\pm=\sqrt{2}(1\pm\phi)/(1+\phi^2)$. The spin 1/2 operators
${\bf S}_{L,R}$ represent the two electron spins on the
DD~\cite{Loss_99}. We arrive at the effective Hamiltonian
\begin{eqnarray}\label{H_K}
H&=&H_{l}+K{\bf S}_L\cdotp{\bf S}_R+\Delta H\,,\\
\Delta H&=&\sum_{\alpha\alpha'}\left(J_{\alpha\alpha'}
{\bf s_{\alpha\alpha'}}\cdotp{\bf S}_+-
V_{\alpha\alpha'}\rho_{\alpha\alpha'}{\bf S}_L\cdotp{\bf S}_R\right.\nonumber\\
&&\left.+I^{+}_{\alpha\alpha'}{\bf s_{\alpha\alpha'}}\cdotp{\bf S}_-
+2iI^{-}_{\alpha\alpha'}{\bf s_{\alpha\alpha'}}\cdotp{\bf T}\right),
\label{DH}
\end{eqnarray}
where 
${\bf s}_{\alpha\alpha'}=
\sum_{kk'\sigma\sigma'}c_{\alpha
k\sigma}^{\dag}(\vec{\bf\sigma}_{\sigma\sigma'}/2)c_{\alpha'k'\sigma'}$,  and 
$\rho_{\alpha\alpha'}=\sum_{kk'\sigma}c_{\alpha k\sigma}^{\dag}c_{\alpha'k'\sigma}$.
In (\ref{H_K}) the bare constants $J,V,I^\pm$ are defined
at the energy cutoff
$D_0\simeq\hbar\omega_0$,
\begin{eqnarray}\label{bare}
&&J=\frac{2}{E_C}V^+,\;\;\;\;\;
V=\frac{\phi_+\phi_-}{2E_C}V^-,\;\;\;\;\;
I^{\pm}=\frac{\phi_{\pm}}{E_C}W^{\pm},
\end{eqnarray}
with matrix elements
$V^{\pm}_{\alpha\alpha'}=t^{*}_{\alpha',+}t_{\alpha,+}\pm
t^{*}_{\alpha',-}t_{\alpha,-}$,
$W^{\pm}_{\alpha\alpha'}=t^{*}_{\alpha',-}t_{\alpha,+}\pm
t^{*}_{\alpha',+}t_{\alpha,-}$, and $E_C=2E_+E_-/(E_++E_-)$. Here,
$E_{\pm}=E(M\pm 1)-E(M)\mp\mu$ is the
CB addition/extraction energy, with $E(M)$ being the energy of the
DD with $M$ electrons, and $\mu=(\mu_L+\mu_R)/2$. We calculate
$E(M)$ for $M=1,2,3$ within the HM method.

Attaching leads to the DD results in a shift of the DD spectrum
such that $K=K_0+\delta K_{SW}$, where
\begin{equation}
K_0={\rm v}-\frac{U_{H}}{2}+\frac{1}{2}\sqrt{U_{H}^2+16t_{H}^2}
\label{K_0}
\end{equation} 
is the exchange interaction of the isolated DD, with ${\rm v}$ accounting 
for long range Coulomb effects\cite{Loss_99}, and $\delta K_{SW}=2\nu
D_0(E_C/E_-){\mbox{Tr}}V <0$ stems from the Schrieffer-Wolff
transformation.  We next derive the RG equations for the coupling
constants in (\ref{DH}) by applying the ``poor man's'' scaling
approach\cite{Anderson}. In matrix form the RG equations
read,
\begin{eqnarray}\label{scaling_total}
&&\dot{J}=J^2+(I^+)^2-(I^-)^2\;,\;\;\dot{V}=2[I^-,I^+],
\nonumber\\
&&\dot{I}^{\pm}=\{J,I^{\pm}\}+[V,I^{\mp}]\, ,
\end{eqnarray}
where the dot denotes $d/d(\nu{\cal L})$. Here $\nu$ is the
density of states per spin  in the leads, and ${\cal
L}=\ln{(D_0/D)}$, with $D$ being the scaled cutoff. Eqs.
(\ref{scaling_total}) are valid for $D\gg K,\Delta\mu, T_0$ (for $T_0$ see
below). The set of 16 equations (\ref{scaling_total}) can be
reduced to 9 equations for our case of $t_{\alpha n}$ being real\cite{note1}.
Solving (\ref{scaling_total}) numerically we find the
characteristic energy scale of the problem,
$T_0=D_0\exp(-\gamma/\nu J_0)$,
where $J_0=(t_L^2+t_R^2)/E_C$, and $\gamma\leq 0.5$ is a
non-universal number, determined by the ratios of the coupling 
constants (\ref{bare}) to $J_0$, and thus depends on the internal
features of the DD. We note that at $t_R=0$ we find $\gamma=(1-S^2)/4$, 
i.e. independent of the Coulomb interaction.
Also, in the limit of separated dots, $t_0=0$,  Eqs.~(\ref{scaling_total}) 
decouple into two RG equations,  determining the Kondo temperatures for 
two 1/2 spins\cite{Anderson}, 
$T_{K(L,R)}^0=D_0\exp(-E_C/4t_{L,R}^2)$.

We derived Eqs. (\ref{scaling_total}) assuming a singlet-triplet
degeneracy (STD) of the DD, 
{\em i.e.} $K=0$. However, the RG procedure generates terms that 
also renormalize the exchange $K$ 
between the DD spins. 
Indeed, we find
\begin{equation}\label{K}
\dot{K}=2\ln(2)D\nu^2{\mbox{Tr}}\left\{V^2+(I^+)^2-(I^-)^2-J^2\right\} .
\end{equation}
The RG flow (\ref{K}) resembles the generation of the RKKY 
interaction in the two-impurity 
Kondo model~\cite{Jones}. It follows from (\ref{K}), (\ref{scaling_total}) 
that, at low $T\sim T_0$, the renormalized $K=K(B,D)$ strongly depends on both 
$D$ and $B$ at the STD point $B_*$ with $K(B_*,D)=0$, see right inset of Fig.~3. 
This suggests that, (i)  $B_*$ shifts towards the triplet 
side as $T$ is lowered down to $T_0$, 
(ii) the energy scale of the problem strongly depends on $K_0$, 
which presumably implies a rather sharp crossover between a spin 1 Kondo 
regime on the triplet side and a locked singlet of ${\bf S}_L$, ${\bf S}_R$ 
({\em i.e.} no Kondo effect) on the singlet side. 

We  perform a non-equilibrium calculation of the current through the DD
up to the second order in the perturbation (\ref{DH}).
The current consists of an elastic and inelastic
component, ${\cal I}_K={\cal I}_{\mbox{\scriptsize\em el}}+
{\cal I}_{\mbox{\scriptsize\em inel}}$,
\begin{equation}\label{I_0}
{\cal I}_{\mbox{\scriptsize\em el}}=\frac{e}{\hbar}\pi\nu^2
\left[J_{LR}^2\left\langle{\bf S}_+^2\right\rangle
+V_{LR}^2\left(\frac{9}{4}
-\left\langle{\bf S}_+^2\right\rangle\right)\right]\Delta\mu\;,
\end{equation}
\begin{equation}\label{I_1}
{\cal I}_{\mbox{\scriptsize\em inel}}=\frac{e}{\hbar\beta}\pi\nu^2
(I_{LR}^-)^2g(\beta K)f(\beta K,\beta\Delta\mu)\;,
\end{equation}
where $\beta=1/T$ (with Boltzman constant set to one),
$g(u)=(e^u+1)/(e^u/3+1)$ accounts for the degeneracy 
of the excited level,
$\langle{\bf S}_+^2\rangle=6/(e^{\beta K}+3)$, and
\begin{equation}\label{fuv}
f(u,v)=\frac{u\tanh(u/2)\sinh(v)+v[1-\cosh(v)]}
{\cosh(u)-\cosh(v)}\;.
\end{equation}
Eqs.~(\ref{I_0}), (\ref{I_1}), together with the RG equations
(\ref{scaling_total}),  describe the renormalization of 
the current ${\cal I}$, {\em i.~e.} 
${\cal I}\propto{\cal I}_K(\ln(D_0/T))$. 
However, not all 
terms were retained in (\ref{H_K}), but only those which become
renormalized. 
To give a complete description of ${\cal I}$ at temperatures 
$T_K\ll T\leq D_0$, we perform a systematic 
calculation of the cotunneling current using the technique developed 
in\cite{Loss_noise}. In the cotunneling regime, including level shifts but 
neglecting heating effects\cite{Loss_noise},  we match the
two currents and obtain for the total  current (after lengthy calculation)
${\cal I}({\cal L})={\cal I}_{K}({\cal L})+\delta{\cal I}(0)$, with
\begin{eqnarray}
&&\delta{\cal I}=\frac{e}{\hbar}\pi\nu^2
\left[
V_{LR}^2\left(\frac{7}{4}-
\left\langle{\bf
S}_+^2\right\rangle\right)+J_{LR}^2
\left(\frac{E_--E_+}{E_-+E_+}\right)^2-\right.\nonumber\\ &&\left.
2V_{LR}J_{LR}\frac{E_--E_+}{E_-+E_+}\left(2-\left\langle{\bf S}_+^2
\right\rangle\right)
\right]
\Delta\mu\;.
\end{eqnarray}
In the cotunneling regime, the differential conductance
$G=ed{\cal I}/d\Delta\mu$ shows a step at $\Delta\mu=\pm K$,
for  $T<|K|$. This step is due to the inelastic
current (\ref{I_1}) contributing for $|\Delta\mu|\geq|K|$.
The step height of $G$ is different for
 the ground state being a singlet
or triplet; the 
ratio of heights is given by  $g(K/T)$.
As $T$ is lowered, Kondo correlations
are expected to develop at $\Delta\mu=\pm K$, with a Kondo peak arising at the
step of $G$. Kondo correlations at finite bias are not the subject of the
present paper, but we would like to point out that the step in $G$ can be
used for determining the ground state of the DD and measuring  $K$.

\begin{figure}
\narrowtext
{\epsfxsize=8.7cm
\centerline{
%\rotatebox{-90}
{\epsfbox{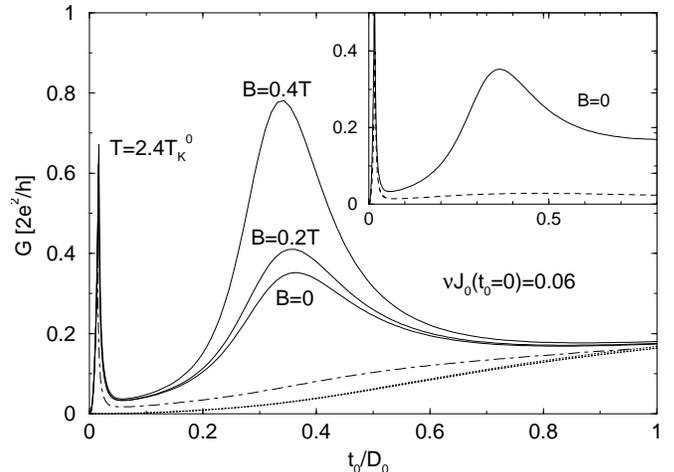}}}}
\vspace{-18pt}\caption{Linear conductance $G(t_0)$ at different $B$'s.
Dotted lines: cotunneling contributions. 
Dot-dashed line: $G$ vs $t_0$ at $B=0.4T$ without the long range part of the 
Coulomb interaction, i.e. for ${\rm v}=0$ in (\ref{K_0}).
For definiteness, we keep $E_+=E_-$ (middle of
CB valley)  by  adjusting $V_g$ when varying $t_0$, and  set  $t_L=t_R$ in 
Figs.~2-4. 
Inset: Comparison of full  (solid
line)  with chain (dashed line)
model Hamiltonian.}
\label{fig2}
\end{figure}

We turn now to a discussion of the linear conductance, $G=G(\Delta\mu=0)$. 
First, we consider the case $B=0$. At small 
$t_0$ the RG growth of $K$ is weak (due to near cancellation of the
trace terms in (\ref{K})), and 
a limit is reached where each  spin is strongly  coupled with a lead electron.
The Kondo effect of the DD, in this case, consists of two independent spin 
$1/2$ Kondo effects for each of the dots separately.
At large $t_0$, the RG correction
to $K$ can be neglected because of a large value of $K_0$.
The two dot spins are locked into a singlet state in this case,
and the lead electrons feel only the potential scattering, slightly enhanced
by the RG flow (\ref{scaling_total}) which terminates at $D=K$.
At intermediate $t_0$, such that $K\sim T_K$,  the exchange interaction
(between the dot spins) and the spin 1/2 Kondo effect compete, and a crossover
between the two regimes  occurs. At this crossover, the
renormalization of K is comparable with the Kondo temperature $T_K$; 
each of the dot spins couples to both leads. 
The conductance versus $t_0$ shows a peak at $t_0\sim\sqrt{U_HT_K}$ 
(see Fig.~2). A second peak at larger $t_0$ emerges 
due to the interplay between Kondo correlations and
long range Coulomb interaction, see dot-dashed line in
Fig.~2. 
This striking sensitivity of the second peak on the long range Coulomb 
interaction provides a way of studying the screening effects in coupled 
quantum dots.
As a check we have taken the Hubbard limit (chain Hamiltonian) of our model, 
and 

\begin{figure}
\narrowtext
{\epsfxsize=8.8cm
\centerline{
%\rotatebox{-90}
{\epsfbox{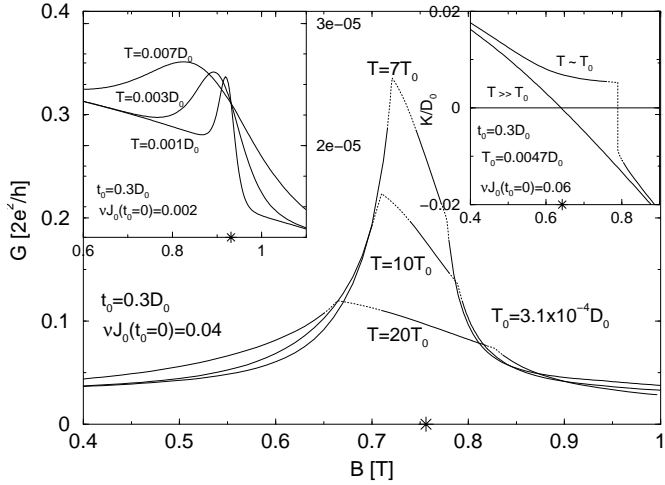}}}}
\caption{Linear $G$ vs $B$ around  $K=0$. The kinks in the
dotted-line regions are an artifact of our two stage
RG procedure in the crossover regime and will be smoothened 
in an exact treatment.
The stars denote $B_*$ with $K(B_*,D=D_0)=0$. 
Left inset: Cotunneling conductance vs $B$ at small coupling to the leads
where Kondo correlations become negligible. 
Right inset: Renormalized $K$ vs $B$, showing  a sharp 
singlet-triplet transition at low 
$T (\sim T_0)$.}
\label{fig3}
\end{figure}

\hspace{-12pt}
found good agreement with the exact NRG calculations of Ref.\cite{Izumida}.
In inset of Fig.~2 we compare
the two models to illustrate the importance of 
the long range Coulomb interaction. 
For small $B$ (with $K>0$), we find that the second peak in $G(t_0)$ 
is very sensitive to $B$, see Fig.~2. 

\begin{figure}
\narrowtext
{\epsfxsize=8.8cm
\vspace{-12pt}
\centerline{
%\rotatebox{-90}
{\epsfbox{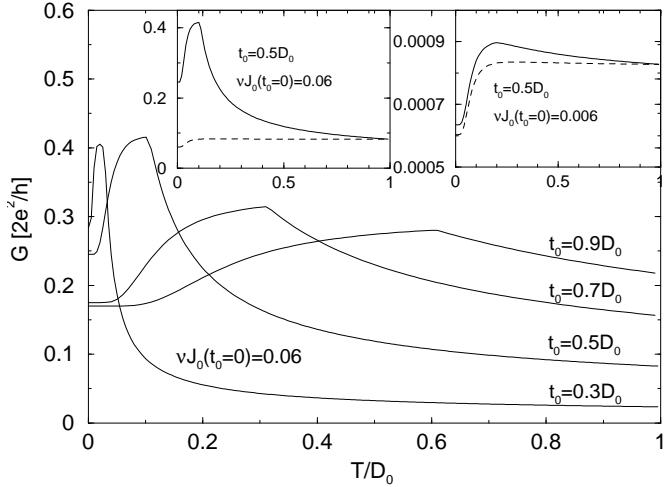}}}}
\caption{
Linear $G$ vs $T$ for different $t_0$. Insets:
Comparison of cotunneling (dashed line) and renormalized (solid line) calculations for
different  couplings to the leads.}
\label{fig4}
\end{figure}

At larger $B$, a singlet-triplet transition occurs\cite{Loss_99}.
Around this point $K=0$, the  RG flow (\ref{scaling_total})
terminates at $D=T>|K|$, which leads to a striking RG enhancement of the
peak in the cotunneling 
conductance, see Fig.~3 and left inset.
In the singlet regime, for $K>T$, 
the RG flow terminates at $D=K$, and there is no Kondo limit.
Physically, this means that
the DD  spins are locked into a singlet with exchange energy
$K>0$. 
In the triplet regime, for $-K>T$, we apply a two stage RG
procedure\cite{Hewson} and find a  $K$-dependent energy scale for
the  spin 1 Kondo effect, very much like as in single
dots~\cite{Glazman,Nazarov}. 
At low $T$, the renormalized  $K$ shows a striking
dependence on $B$ around $K=0$, see right inset of Fig.~3. Our perturbative RG
thus suggests a sharp transition between the two regimes,
with a rapidly vanishing spin 1 Kondo temperature when approaching the singlet side.
Finally, in Fig.~4 we plot the conductance vs $T$. We see that again
the Kondo correlations lead to a pronounced peak in G which occurs at
$T=K$, thus providing a further way to access the exchange
interaction $K$.

{\it Acknowledgments.} We thank G.~Burkard, L. Glazman, and
M. Pustilnik for 
discussions. We acknowledge support from the Swiss NSF and DARPA.


\vspace{-10pt}
\begin{references}
\vspace{-40pt}
\bibitem{Kouwenhoven}
L.P. Kouwenhoven, G. Sch\"on, L.L. Sohn,
{\it Mesoscopic Electron Transport},
NATO ASI Series E, Vol.\ 345,
Kluwer Academic Publishers (1997).

\bibitem{Raikh} 
L.I.~Glazman, M.E.~Raikh, 
JETP Lett. {\bf 47}, 452 (1988);
T.K.~Ng, P.A.~Lee, Phys.\ Rev.\ Lett. {\bf 61} 1768 (1988).


\bibitem{GG} 
D.~Goldhaber-Gordon, {\em et al.} Nature (London) {\bf 391},
156 (1998).

\bibitem{Cronenwett}
S.M.~Cronenwett, T.H.~Oosterkamp, L.P.~Kouwenhoven, Science {\bf 281},
540 (1998). 

\bibitem{Schmid}
J.~Schmid, {\em et al.} Physica B {\bf 256-258}, 182 (1998).

\bibitem{Cobden}
T.W.~Odom, {\em et al.} Science {\bf 290}, 1549 (2000);
J.~Nygard, D.H.~Cobden, P.E.~Lindelof, Nature (London) {\bf 408},
342 (2000).

\bibitem{Sasaki} 
S.~Tarucha, {\em et al.} Phys.\ Rev.\ Lett.\ {\bf 84}, 2485 (2000);
S.~Sasaki, {\em et al.} Nature (London) {\bf 405}, 764 (2000);
J.~Schmid, {\em et al.} Phys.\ Rev.\ Lett.\ {\bf 84}, 5824 (2000).

\bibitem{Glazman}
M.~Pustilnik, L.I.~Glazman, Phys.\ Rev.\ Lett.\ {\bf 85}, 2993
(2000);
Phys.\ Rev.\ B {\bf 64}, 045328 (2001).

\bibitem{Nazarov}
M.~Eto, Y.~Nazarov, Phys.\ Rev.\ Lett.\ {\bf 85}, 1306 (2000);
Phys.\ Rev.\ B {\bf 64}, 085322 (2001).

\bibitem{Hofstetter} W. Hofstetter, H. Schoeller, cond-mat/0108359.

\bibitem{Eto}
T.~Aono, M.~Eto, Phys.\ Rev.\ B {\bf 63} 125327 (2001);
A.~Georges, Y.~Meir, Phys.\ Rev.\ Lett.\ {\bf 82}, 3508 (1999).

\bibitem{Izumida}
W.~Izumida, O.~Sakai, Phys.~Rev.~B {\bf 62}, 10260 (2000).

\bibitem{Loss97}
D. Loss, D.P. DiVincenzo,
Phys.\ Rev.\ A {\bf 57}, 120 (1998).

\bibitem{Kikkawa}
J.M. Kikkawa, D.D. Awschalom, Phys.\ Rev.\ Lett.\ {\bf 80},
4313 (1998).

\bibitem{Loss_99}
G.~Burkard, D.~Loss, D.P.~DiVincenzo,
Phys. Rev. B {\bf 59} 2070 (1999).

\bibitem{2dots_exper}
F.R.~Waugh {\it et al.},
Phys.\ Rev.\ Lett.\ {\bf 75} 705 (1995).

\bibitem{SW}
J.R.~Schrieffer, P.A.~Wolff,
Phys.\ Rev.\ {\bf 149}, 491 (1966).

\bibitem{Anderson}
P.W.~Anderson, J.~Phys.~C {\bf 3}, 2436 (1970).

\bibitem{note1}
Eqs.~(\ref{scaling_total}) hold for arbitrary $t_{\alpha n}$ and can 
also describe a parallel (Aharonov-Bohm) geometry of the DD.

\bibitem{Jones}
B.A.~Jones, C.M.~Varma, Phys.\ Rev.\ Lett.\ {\bf 58},
843 (1987). 

\bibitem{Loss_noise}
E.V.~Sukhorukov, G.~Burkard, D.~Loss,
Phys. Rev. B {\bf 63} 125315 (2001).

\bibitem{Hewson}
A.C.~Hewson, {\it The Kondo Problem to Heavy Fermions}, 
(Cambridge Univ. Press, 1997).

\end{references}
\end{document}